\documentclass[aps,twocolumn,showpacs,superscriptaddress,amsmath,amssymb]{revtex4-1}
\usepackage{graphicx}
\usepackage[utf8]{inputenc}
\usepackage[pdftex]{hyperref}
\usepackage{multirow}
\usepackage{color}
\usepackage{xspace}
\usepackage{gensymb}  
\usepackage{epstopdf}

\DeclareGraphicsExtensions{.jpg, .png , .pdf, .bmp}

\begin{document}

\title{Growth Modes and Chiral Selectivity of Single-Walled Carbon Nanotubes}

\author{Maoshuai He}
\affiliation{Key Laboratory of Eco-Chemical Engineering, Ministry of Education, College of Chemistry and Molecular Engineering, Qingdao University of Science and Technology, Qingdao 266042, China}
\affiliation{School of Materials Science and Engineering, Shandong University of Science and Technology, 266590, Qingdao, People's Republic of China.}

\author{Yann Magnin}
\affiliation{Centre Interdisciplinaire de Nanoscience de Marseille, Aix-Marseille University and CNRS, Campus de Luminy, Case 913, F-13288, Marseille, France.}

\author{Hua Jiang}
\affiliation{Department of Applied Physics, Aalto University School of Science, P.O. Box 15100, FI-00076 Aalto, Finland.}

\author{Hakim Amara}
\affiliation{Laboratoire d'Etudes des Microstructures, ONERA-CNRS, BP 72, 92322 Châtillon Cedex, France.} 

\author{Esko I. Kauppinen}
\affiliation{Department of Applied Physics, Aalto University School of Science, P.O. Box 15100, FI-00076 Aalto, Finland.}

\author{Annick Loiseau}
\affiliation{Laboratoire d'Etudes des Microstructures, ONERA-CNRS, BP 72, 92322 Châtillon Cedex, France.} 

\author{Christophe Bichara}
\affiliation{Centre Interdisciplinaire de Nanoscience de Marseille, Aix-Marseille University and CNRS, Campus de Luminy, Case 913, F-13288, Marseille, France.}

\date{\today}

\begin{abstract}

Chemical vapor deposition synthesis of single-walled carbon nanotubes experiments, using Fe catalyst, and alternating methane and carbon monoxide as carbon feedstocks, lead to the reversible formation of junctions between tubes of different diameters. Combined with an atomistic modeling of the tube / catalyst interface, this shows that the ratio of  diameters of the tube and its seeding particle, denoting the growth mode, depends on the carbon fraction inside the catalyst. With carbon monoxide, nanoparticles are strongly carbon enriched, and tend to dewet the tube, in a perpendicular growth mode. Cross-checking our results with available reports from the literature of the last decade strongly suggests that these latter conditions should favor the near armchair chiral selectivity observed empirically. 

\end{abstract}

\maketitle

\section{Introduction}

Carbon Nanotubes (CNTs), among which single-walled ones (SWNTs) are most promising, have now reached a high level of maturity. A number of  applications based on SWNTs outstanding electronic properties, such as transparent conductive films \cite{Kaskela10} or electronic components\cite{Brady2016} have been proposed, and will be further developed if efficient and selective synthesis techniques are available for large scale production. Catalytic chemical vapor deposition (CVD) is the method of choice to synthesize SWNTs, because it can be upscaled to industrial requirements. In this process, carbon bearing molecules decompose on a nanosized catalyst particle,  eventually giving rise to a tube that keeps growing until some event causes the growth to stop\cite{Jourdain13}. All this takes place at high temperature (700-1400 K), in a complex ambient, with catalyst nanoparticles either bound to a substrate, or floating at the tip of the growing tube. Quite amazingly, more than 25 years after their discovery\cite{Iijima91} and first synthesis\cite{Iijima93}, a recent review \cite{Liu2016} still stresses the lack of complete understanding of the SWNT growth mechanisms. The control of the tubes' properties by selective synthesis, clearly a major bottleneck towards applications, is consequently still elusive. 

A pair of  indices $(n, m)$ defines the tube structure, - its "chirality", or more properly, its "helicity" -, and electronic properties. Starting with the pioneering work introducing the CoMoCAT catalyst \cite{Bachilo2003}, a number of catalytic systems have been proposed for chiral-selective SWNT growth. Since then, when some degree of chiral selectivity is reported in  CVD growth experiments, it usually  corresponds to a higher occurrence of large chiral-angle SWNTs (i.e. near-armchair) in the produced nanotube samples\cite{Bachilo2003, Wang2013, Miyauchi2004, Wang2007, Li2007, Kajiwara2009, Chiang09, He2010, Loebick2010, Fouquet2012, He2014}.  Significant progress towards $(n,m)$ selective synthesis using solid state catalyst has recently been reported\cite{Yang14, Zhang2017}. \ Unfortunately, no fully satisfactory model is available to explain these results, especially regarding the role of the catalyst and its influence on selectivity. This limited understanding of the growth mechanisms inhibits progress towards in-growth selectivity. Available theoretical modeling \cite{Ding09, Kim2012} and experimental results \cite{Rao2012, Liu2013} suggest that the observed chiral distributions favoring near armchair SWNTs results from their faster growth rates. Further developments \citep{Artyukhov2014a} try to incorporate interfacial energy contributions, and address the atomic scale incorporation of carbon atoms in the tube, but fail at giving realistic tube chiral distributions. In this context, recent transmission electron microscopy (TEM) and electron diffraction (ED) analysis of a statistical number of SWNTs showed no correlation between the length and the chiral angle of tubes grown by CVD with Fe catalyst and CH$_4$ feedstock\cite{He2017}, thus casting doubts on the validity of these models.

In this paper, we show experimentally that a rational control of the tube growth mode, introduced by Fiawoo \emph{et al.}\cite{Fiawoo12}, is now possible. These modes were empirically characterized by the aspect ratio $R_d = D_{CNT} / D_{NP}$ between the diameter of the tube ($ D_{CNT}$) and that of the nanoparticle (NP) from which it grew ($D_{NP}$), from which perpendicular ($R_d < 0.75$) and tangential ($R_d \geq 0.75$) growth modes can be distinguished.Monte Carlo computer simulations on these nanosized objects meet previous knowledge of metallurgists relevant to steel carburization, to relate these modes to the way carbon is dissolved in the catalyst during growth. Systematic TEM analyses of the sizes of the SWNTs and the NPs from which they grew, as well as tubes' helicity, indicate that the sometimes reported near armchair selectivity only stands for tubes grown in perpendicular mode. Our results strongly suggest that a near armchair selectivity can be obtained under appropriate conditions by tuning the carburization level. This can be achieved by selecting the appropriate metal and carbon precursor, or acting on pressure and temperature, hence dewetting between the SWNT and the carbon-rich catalyst to favor the perpendicular mode.  

\section{Methods}

The Fe catalyst was prepared by hydrolysis of anhydrous ferric chloride in boiling water\cite{He04}. The prepared Fe colloid was diluted in ethanol and dispersed onto silicon nitride grid or Au grid covered by SiO$_x$ membrane. Supported Fe particles were calcined (1073 K, 2 h) in air to remove organic residual and loaded into the center of a horizontal quartz tube CVD reactor. After heating to 1073 K under the protection of helium and being stabilized, the helium was switched off and the carbon source was introduced by alternating cycles of 1 min CH$_4$ and 1 min CO. Control experiments, with only CH$_4$ or CO were also performed. The morphologies of catalyst particles and carbon nanotubes were characterized by JEOL-2200FS FEG TEM/STEM. Structural assignments of the SWNTs were performed by analyzing their nanobeam patterns. On the basis of intrinsic layer line spacing strategy\cite{Jiang07}, the diffraction pattern was indexed and the chiral indices (n, m) of the SWNT were determined.

\section{Controlling SWNT growth modes in Fe-based CVD synthesis.}

\begin{figure}[htb!]
\begin{center}
\includegraphics[width=1\linewidth]{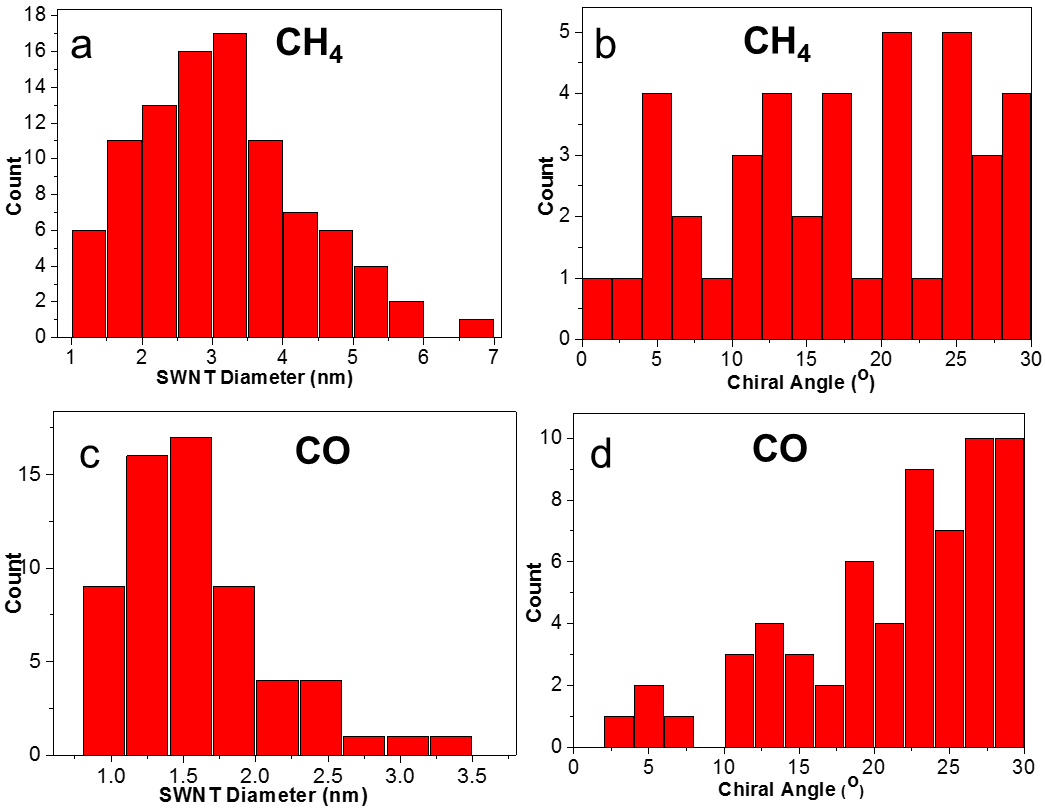}
 \caption{(Color online) (a) Diameter and (b) chiral angle distributions of SWNTs grown on Fe nanoparticles at 1073 K using CH$_4$ as the carbon source. (c) Diameter and (d) chiral angle distributions of SWNTs grown on Fe nanoparticles at 1073 K using CO as the carbon source. }
  \label{fig:Fig1}
\end{center}
\end{figure}

Tubes were grown at 1073 K using two different CVD reactors. The first one is a horizontal furnace, described in \citep{He2017}, that enables surface bound growth of SWNTs using Fe as a catalyst, and CO or CH$_4$ as carbon feedstocks. Catalyst preparation and experimental setups and protocols are described in section Methods. After growth, TEM and nanobeam electron diffraction were used to thoroughly characterize a statistically significant amount of SWNTs. Results of this analysis are presented in Fig.\ref{fig:Fig1}. Tubes grown with CH$_4$ display a broad diameter distribution, with and average around 3 nm, and no chiral selectivity. The diameter distribution of tubes grown with CO is narrower (average diameter around 1.5 nm), and a clear preference towards large chiral angles can be seen.

\begin{figure}[htb!]
\begin{center}
\includegraphics[width=0.80\linewidth]{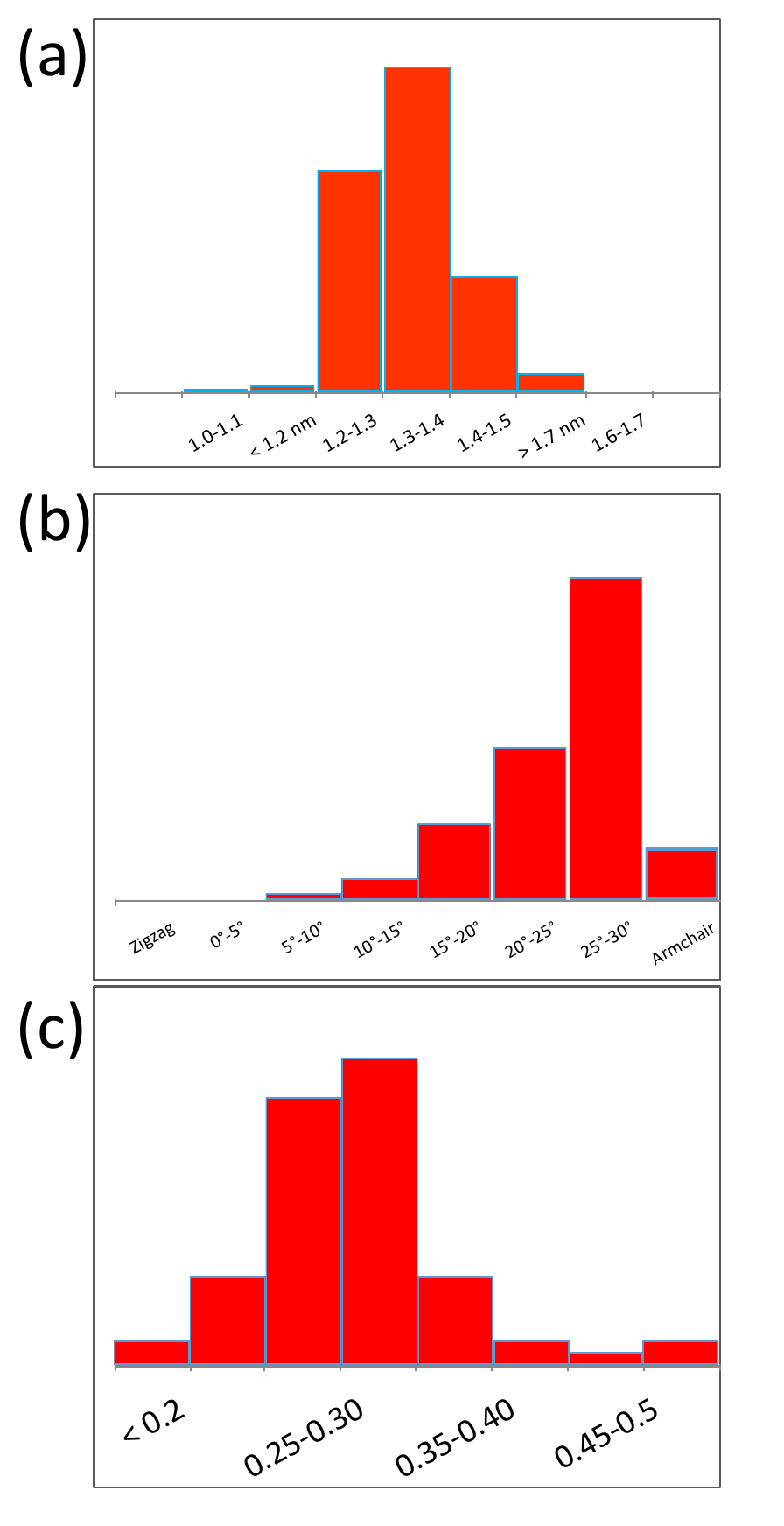}
 \caption{(Color online) (a) Diameter, and (b) chiral angle distribution measured by nanobeam electron diffraction on a batch of 166 single-walled carbon nanotubes grown by floating CVD, using ferrocene and CO feedstocks. Tube diameters are sharply peaked around 1.35 nm, and a trend towards near armchair helicities is clearly visible. It is worthy noting about 8\% of observed nanotubes have armchair chirality.  (c) Distribution of SWNT/NP diameter ratios measured for active NPs that gave rise to a SWNT, indicating a perpendicular growth mode.}
  \label{fig:Fig2}
\end{center}
\end{figure}

This preference for near armchair selectivity is even more pronounced in experiments using a floating catalyst CVD setup \citep{Moisala06}. Another series of tubes were grown at 1073 K, using ferrocene (forming Fe catalyst NPs within the CVD reactor), and CO as a feedstock, and TEM-based statistical investigations were performed in a similar way. Results are displayed in Fig.\ref{fig:Fig2}. Statistics over 166 measured tubes reveal a strong preference towards near armchair chiral angles and diameters around 1.3 nm. This tube diameter distribution is significantly sharper than for surface bound CVD, because of a narrower NP size distribution. The aspect ratio $R_d$ distribution, displayed in Fig.\ref{fig:Fig2}c, peaked around 0.35, clearly indicates that, statistically, tubes are grown in a perpendicular mode.\\
Similar results were already presented, in particular by Lolli \textit{et al.}\cite{Lolli06} and He \textit{et al.}\citep{He2013a}, but the present data go one step further by clearly indicating that a close to armchair selectivity is only obtained for tubes grown in perpendicular mode, with CO, and not those grown in tangential mode, with CH$_4$.
 
\begin{figure}[htb!]
\begin{center}
\includegraphics[width=0.75\linewidth]{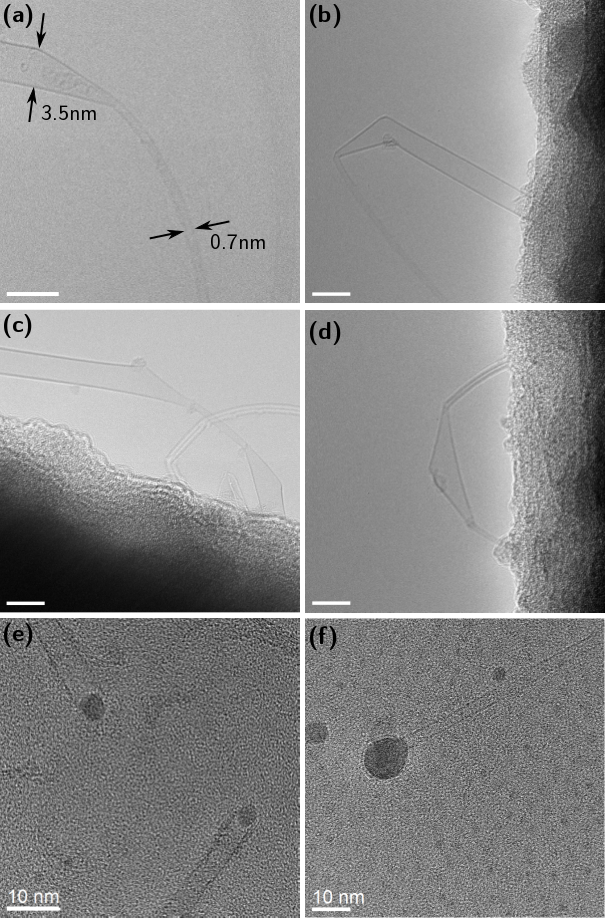}
  \caption{(a)-(d) TEM images of SWNT intramolecular junctions formed by alternating CO and CH$_4$ during CVD growth processes at 1073 K. Scale bars in all images are 5 nm. TEM images of SWNTs grown by (e) tangential mode by CH$_4$ and (f) perpendicular mode by CO at 1073 K.}
  \label{fig:Fig3}
\end{center}
\end{figure}

A second important step is to show that growth modes can be \textit{reversibly} switched by alternating CO and CH$_4$ feedstocks, thus ruling out any change of the catalyst as a cause for the SWNT diameter change. To this end, surface bound CVD experiments with Fe at 1073 K were performed, alternating CO and CH$_4$ feedstocks (approximately 1 min each, up to 15 switches). Using previous data as control experiments, we could check that diameters obtained with each feedstock were in the same range, using either a single feedstock, or alternating both. Figure \ref{fig:Fig3} presents TEM images of tubes obtained in such a way. Clearly, each SWNT consists of two or more segments with different diameters where the percentage of tubes with altered diameters is about 20\% by changing carbon sources on the basis of TEM characterizations. For most SWNT junctions reported in previous studies, the diameter change along the tubes\cite{Ouyang01, Yao07, Thurakitseree13} ranges between 10 and 100 \%. In our experiments, statistics show that the diameter ratios of thick to thin segments for SWNT junctions range from 1.4 to 6.6, with a mean of 3.6, i.e. 260 \% diameter difference. Because of the unknown incubation time and the possible tube nucleation at any time, the actual growth time to form each SWNT segment is not exactly known, making statistics on the segments' length irrelevant. 

It is important to notice that the diameter alternation along SWNT junctions is reversible (Figure \ref{fig:Fig3}(a-d), corresponding to the reversible alternation of carbon  sources during CVD reaction. This clearly rules out irreversible changes of the nanoparticle size, induced by either Ostwald ripening or coalescence, as suggested by Lolli \textit{et al.}\cite{Lolli06}, as the cause of the diameter changes when using different carbon feedstocks.\\

These experiments, demonstrating a reversible diameter change of SWNTs by alternating feedstocks, leading to decomposition reactions with different thermochemical characteristics (exothermic for CO, endothermic for CH$_4$\cite{Lolli06}) call for an interpretation relying on a thermodynamic basis. Because the reaction of carbon rich precursors on bulk Fe is a key step for steel carburization, thermodynamics and kinetics of carburization of Fe, using either CH$_4$ or CO, have been extensively investigated for macroscopic systems. Such surface treatments can considerably improve mechanical and chemical properties of steels by controlling the carbon concentration inside the material. As determined experimentally, the rate constant for Fe carburization, resulting from CH$_4$ decomposition is low, equal to 1.9$\times$10$^{-6} mol / (cm^2.s.bar)$ at 1193 K. The rate of CO carburization is two orders of magnitude larger, at 1.5$\times$10$^{-4} mol / (cm^2.s.bar)$\cite{Grabke75}. Similar results were also reported by Hosmani \emph{et al.}\cite{Hosmani14}. Consequently the carbon concentration in the NP delivered by CO disproportionation (Boudouard reaction) should be much higher than that produced by CH$_4$ decomposition. Such differences, easy to measure for large systems, are hardly evidenced experimentally at the nanometric scale of the CVD process. We thus turn to atomistic computer simulations that are well suited for investigations at such small sizes.\\

\section{Atomistic computer simulation of the tube/catalyst interface}

\begin{figure}[htb!]
\begin{center}
\includegraphics[width=0.75\linewidth]{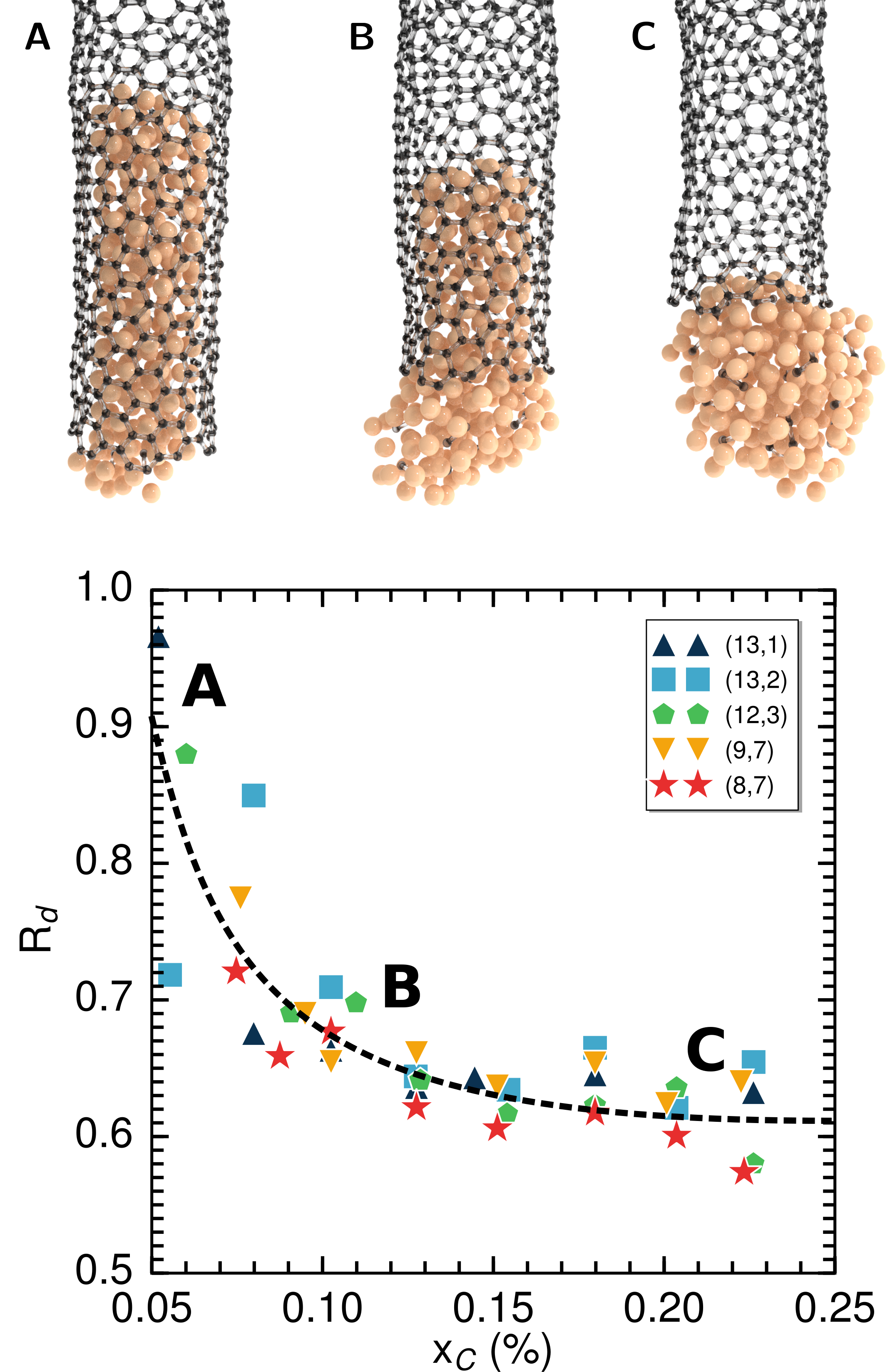}
  \caption{(Color online) Top: Three snapshots (A, B, C) of typical structures of a SWNT at the tip of a Ni$_{219}$ nanoparticle, with different fractions of carbon dissolved in, as obtained from Monte Carlo simulations at 1400 K. From left to right : $x_c = 0.04, 0.10, 0.23$. Bottom: Tube versus NP diameter aspect ratio ($R_d$), resulting from Monte Carlo simulations (symbols) for tubes with different chiralities and different carbon fractions in the NP ($x_c$). $R_d$ depends strongly on ($x_c$), and also on the tubes' diameter, between 1.0 and 1.1 nm, and chiral angle. These two contributions cannot be disentangled. Configurations (A, B, C) are qualitatively located in the  ($x_c, R_d$) plot.}
  \label{fig:Fig4}
 \end{center} 
\end{figure}

Our previous studies of carbon solubility in nickel nanoparticles \cite{Diarra12a, Diarra12b}, have shown that the wetting angle of a nanoparticle deposited on a graphene layer depends on the fraction of carbon dissolved in it, exhibiting a similar, though more marked trend as the experimental results of Naidich \emph{et al.}\cite{Naidich71} for macroscopic Ni drops on graphite. These results were obtained using  our Tight Binding fourth moment model for Ni-C interactions\cite{Amara09}, implemented in Monte Carlo simulations in  canonical, or grand canonical ensemble\cite{Frenkel02}. The same technique is used here to study the equilibrium shape of nanoparticles with 219 Ni atoms, and fractions of carbon ($x_c$) between 0 and 23\%, located at the tip of SWNTs with different chiralities and diameters between 0.8 and 1.2 nm. Sufficient tube length was allowed for the NP to be possibly entirely sucked in the tube, keeping a distance from both ends larger than the cut-off distance of our energy model. The  different samples were relaxed at 1400 K for a long enough time, in order to reach equilibrium. For a macroscopic system true equilibrium should lead to a dissolution of enough carbon atoms from the tube to reach the solubility limit of the metal particle. Because of the nanometric size of the system, interfacial energy contributions become relatively more important, and modify this picture. \\
Typical results are displayed in Figure \ref{fig:Fig4}, showing that the equilibrium shape depends on the carbon fraction dissolved in the NP. Pure Ni NP tends to be completely sucked inside the tube, while fully carbon saturated ones remain completely outside the tube, with the tube edge as the remaining contact line between tube and nanoparticle. Between these two extremes, intermediate situations are found, with a part of the NP inside the tube, and another part outside. We also note that dissolved carbon atoms tend to remain in the outer part of the NP, while the inner part, in contact with the SWNT wall, remains essentially depleted of carbon. Such a depletion has already been observed experimentally  in the case of a graphene layer deposited on a Ni (111) surface, using \emph{in situ} XPS \cite{Benayad13, Weatherup14} and explained \cite{Weatherup14} on the basis of Density Functional Theory (DFT) and tight binding calculations. Moreover, the stability of tubes attached to NPs which were either pure Ni or Ni with up to 20 \% C has been investigated by performing DFT calculations \cite{Anders2010}. It was found that the adhesion of the tube is stronger with a pure Ni cluster than for a Ni NP containing C. Consequently, as long as the tube diameter remains small enough for dissolved carbon to "feel" the SWNT wall, no carbon enters inside the tube with the Ni atoms. For very narrow tubes (below $\approx$ 0.7 nm), nor do metal atoms enter inside the tube. 

We can further analyze these computer simulation results in relation with our observation of tangential and perpendicular growth modes \cite{Fiawoo12}. Pure metal NPs inserted inside the tube, with a large surface contact with the inner part of the carbon sp$^2$ wall correspond to a tangential situation. Carbon saturated NPs, completely outside the tube, with no contact but the tube edge are so-called "perpendicular", even if the limited NP size used here makes this picture a bit inappropriate. Assuming a spherical shape for the outer part of the NP and an atomic density corresponding to liquid Ni at 1400K, the aspect ratio $R_d$ can be calculated. As shown in Figure \ref{fig:Fig4}, it depends strongly on the carbon fraction in the NP, and weakly on the tubes' chirality. In addition, we note that the gradual change in diameter observed in the TEM images is fully consistent with the fact that, upon switching feedstocks, the gas phase composition in the reactor chamber changes smoothly, and so does the carbon concentration in the NP.

\begin{figure}[htb!]
\begin{center}
\includegraphics[width=1\linewidth]{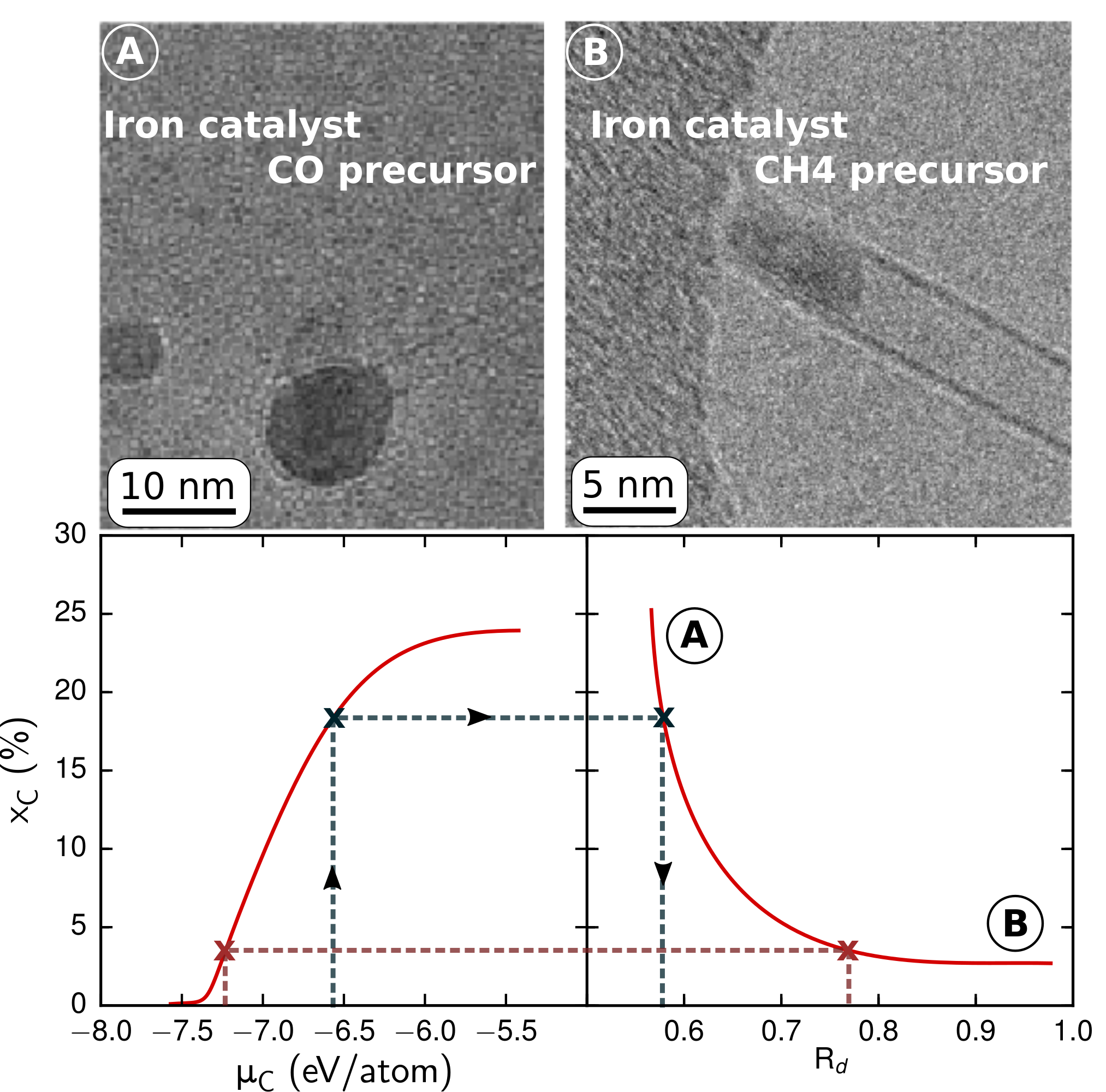}
  \caption{(Color online) Top : TEM images of tubes grown by perpendicular (A) or tangential (B) growth modes. Bottom: Sketch of the relation between the chemical potential of carbon atoms delivered at the surface of the NP ($\mu_{c}$), controlled by the precursor decomposition reaction, and the growth mode $R_d$, via the carbon fraction $x_c$ in the NP. The left curve is a carbon sorption isotherm, as calculated in Diarra \emph{et al.}\cite{Diarra12a}.}
  \label{fig:Fig5}
\end{center}
\end{figure}

This observation sheds a new light on the way to control the growth mode and tube aspect ratio ($R_d$), and on the interpretation of our experiments leading to the formation of nanotube junctions. The carbon concentration in the NP ($x_c$) depends on its chemical potential ($\mu_c$) that is fixed by the thermochemistry of the catalyst decomposition of the NP surface\cite{Snoeck97}. The carbon sorption isotherms calculated in \cite{Diarra12a, Diarra12b} give quantitative estimates of ($x_c$) for different NP structures (fcc or icosahedral) and temperatures. Figure \ref{fig:Fig5} qualitatively connects the growth mode, characterized by $R_d = D_{CNT} / D_{NP}$ and $\mu_c$. Low (more negative) $\mu_c$ support tangential growth, while larger, less negative ones, lead to larger $x_c$, hence perpendicular growth. In CVD experiments, controlled parameters, such as pressure, temperature, the nature of the carbon feedstock or the ambient, can shift $\mu_c$ towards higher or lower values, according to the specific decomposition reaction considered, and hence control $R_d$ and the growth mode. 
In the present experiments, Fe nanoparticles were used as catalyst, while a model for Ni was used in our calculations. However, as shown in\cite{Naidich71}, Ni, Co and Fe behave in the same way regarding the influence of carbon solubility on their wetting properties, that are indeed the key for controlling growth modes. We can thus safely use our calculated results on Ni to interpret the experimental trends observed with Co and Fe. We thus reasonably conclude that CO based growth leads to carbon saturated NPs, hence perpendicular growth, while CH$_4$ yields less C in the NP, favoring aspect ratios  $R_d$ closer to $1$.
\section{Discussion and Conclusion}

SWNT synthesis experiments were performed, using Fe catalyst, and either CH$_4$ or CO as carbon feedstocks, coupled with systematic HR-TEM analysis of the link between the tubes and the NP from which they grew. Electron diffraction analysis of a few hundred tubes allowed for an unambiguous assignment of the tubes $(n,m)$ index. With CO feedstock, tubes were grown in perpendicular mode, and a close to armchair preference was observed, for both supported and floating catalyst CVD, while those with CH$_4$ were closer to tangential mode, and displayed no selectivity. These different behaviors regarding selectivity was already noticed by He \emph{et al.} \cite{He12, He2013a}, but new experiments presented here, alternatively growing tubes with  CH$_4$ or CO indicate that the NP size remains unaffected, while producing different tube diameters. The preference to large chiral angles, assigned here to tubes growing in perpendicular mode, has been explained by a combination of catalyst interface thermodynamics and a kinetic growth theory \cite{Artyukhov2014a}. The calculated chiral distributions are however extremely sharply peaked, as compared to experimental ones, meaning that further developments are still required. On the basis of Monte Carlo simulations, we assign this change of growth mode to changes of the fraction of carbon dissolved in the NP. This understanding opens a way to control the tube/NP diameter ratio, hence tube diameter, an important step forward, since many properties depend on it. 

Indeed, on the basis of our present investigations, and revisiting a number of published results, we suggest that a key to achieve a higher degree of chiral selectivity is to promote a perpendicular growth mode, with an interface between the tube and its seeding particle limited to a line. In the case of catalysts displaying a significant carbon solubility and a moderate melting temperature, such as Fe, Co or Ni, this is achieved by tuning the degree of  carburization of the nanoparticle, via the choice of the appropriate feedstock and growth (P, T) conditions. Under such conditions, a  near armchair selectivity can be obtained. Although not explicitly stated as such, it appears that perpendicular growth mode is also observed or assumed for Co$_{7}$W$_{6}$\cite{Yang14}, as well as WC and Mo$_{2}$C\cite{Zhang2017} that are very selective catalysts. An element to be taken into account to explain these results is that growing tubes in a fully perpendicular growth mode decouples the tube and nanoparticle diameters, thus setting less stringent constraints on the nanoparticle size distribution, and also on the choice of growth conditions. Whatever technique used, it is very difficult or impossible to obtain a very sharply peaked NP size distribution, and to keep it stable under CVD conditions to grow tubes. Moreover, it has been observed that for given  growth (P, T, ambient) condition, only a fraction of the NP size distribution is "activated" to nucleate and grow tubes.

Growing tubes in perpendicular growth modes somewhat alleviates these constraints, while putting more emphasis on the perpendicular contact between the tube and the catalyst. Under such conditions, the choice and the preparation of the catalyst appears as a key step for mastering the tube's structure during the growth. In this context, we should note that new promising nanoparticles synthesis techniques for bimetallic catalysts based on the use of Prussian blue analogs have been developed to produce a wide range of homogenous bimetallic catalyst nanoparticles with controlled stoichiometry and sizes \cite{Castan17}. On the theoretical side, further investigations should be performed to guide the choice of catalysts favoring chiral selectivity.

\section*{Acknowledgments}
Supports from the European Union Seventh Framework Programme (FP7/2007-2013) under grant agreement 604472 (IRENA project) and from the French research funding agency (ANR), under grant 13-BS10-0015-01 (SYNAPSE), are gratefully acknowledged. M. He would also like to acknowledge the Natural Science Foundation of Shandong Province of China (No. ZR2016EMM10) and Scientific Research Foundation of Shandong University of Science and Technology for Recruited Talents (No. 2016RCJJ001).   

\bibliographystyle{apsrev4-1}

\end{document}